# Advances of Artificial Intelligence in Classical and Novel Spectroscopy-Based Approaches for Cancer Diagnostics. A Review

Marina Zajnulina, Dr, Photonic Scientist and Physicist
https://orcid.org/0000-0002-9666-0534
Contact: marina@physik.tu-berlin.de
7.08.2022

**Abstract**
Cancer is one of the leading causes of death worldwide. Fast and safe early-stage, pre- and intra-operative diagnostics can significantly contribute to successful cancer identification and treatment. Artificial intelligence has played an increasing role in the enhancement of cancer diagnostics techniques in the last 15 years. This review covers the advances of artificial intelligence applications in well-established techniques such as MRI and CT. Also, it shows its high potential in combination with optical spectroscopy-based approaches that are under development for mobile, ultra-fast, and low-invasive diagnostics. I will show how spectroscopy-based approaches can reduce the time of tissue preparation for pathological analysis by making thin-slicing or haematoxylin-and-eosin staining obsolete. I will present examples of spectroscopic tools for fast and low-invasive ex- and in-vivo tissue classification for the determination of a tumour and its boundaries. Also, I will discuss that, contrary to MRI and CT, spectroscopic measurements do not require the administration of chemical agents to enhance the quality of cancer imaging which contributes to the development of more secure diagnostic methods. Overall, we will see that the combination of spectroscopy and artificial intelligence constitutes a highly promising and fast-developing field of medical technology that will soon augment available cancer diagnostic methods.

## 1. Introduction
### 1.1 Cancer as the plague of our times

According to the World Health Organisation (WHO), cancer belongs to the leading causes of death around the world and constitutes a large group of diseases that are denoted by the uncontrollable abnormal cell growth of an organ or a tissue. If these cells invade adjoining parts of the body or spread to other organs, we call this process metastasis. Widespread metastases are the primary cause of death from cancer (WHO: Cancer, 2022). According to Global Cancer Statistics, 19.3 million new cancer cases occurred globally in 2020 (Sung et al., 2021). Around 10 million people died from it in the same year. Alone in the USA, around 1.92 million new cancer cases and 610 thousand cancer deaths are projected to happen in 2022 (Siegel et al. 2022). The cancer types leading by the number of new cases are breast, lung, colon and rectum, prostate, skin, and stomach cancers. The deadliest types among them are malignant tumours (cancers) of the lung, colon and rectum, as well as liver (WHO: Cancer, 2022; Wild et al., 2020).

Whereas the number of deaths caused by, for instance, infectious diseases has been constantly declining for the last 60 years due to the improvement of hygiene and medical standards, the number of cancer deaths is increasing due to the growing age of the overall population. Thus, WHO expects the number of new annual cancer cases to exceed 27 million by 2040 (Wild et al., 2020). New developments in cancer diagnostics and treatment can help slow down this trend. In this review, I focus on the advances of Artificial Intelligence (AI) in oncology, a branch of medicine that deals with the prevention, diagnostics, and treatment of cancer. I show how AI can enhance classical diagnostic imaging techniques such as MRI (magnetic resonance imaging) or CT (computed tomography) and provide an overview of novel spectroscopy-based approaches for faster and more secure early-stage as well as intra- and post-operative diagnostics.

## 1.2 Challenges and requirements in cancer diagnostics and treatment

For decades, the diagnosis of cancer and its type has been done using body fluids (blood, plasma, urine, etc.) laboratory tests, imaging techniques including MRI and CT scans, and biopsy analysis done by pathologists using body tissue samples (National Cancer Institute, 2019). Tests of body fluids rely on biochemical analysis and lead to objective results. Analysis of tissue can be less objective. Thus, the pathologic tissue diagnosis is a multi-step process that includes the biopsy, i.e. the operative tissue removal, preparation of the histopathological tissue specimens, analysis of these samples, and the statement of the diagnosis itself. Although these steps are performed by experts in surgery, pathology, oncology, radiology, and other specific fields of medicine, human- or technology-related mistakes and errors can always happen (Pena & Andrade-Filho, 2009). Thus, the biopsy is an invasive operation and, thus, can be dangerous for the patient. The preparation of the histopathological specimens from the resected tissue is a time-consuming process. Formalin-fixed paraffin-embedded tissue specimens used for gold-standard histopathology can take up to several days to prepare. Rapidly frozen, thin-sliced, and stained with haematoxylin-and-eosin tissue samples still require around 30 minutes to be produced (Matsumoto et al., 2019). The time used to prepare the histopathological samples is the time taken from the patient. The specimen analysis under microscopes also suffers from limitations. The field of view of the microscope lens is often smaller than the specimen size. To produce an image from the entire specimen, several smaller images have to be taken and stitched together to a bigger image. It requires pre-processing the initial smaller images to correct for possible aberration, vignetting, and variation in the exposure effects before the actual stitching (Legesse et al., 2015). It is a further step where things can go wrong, and the success of the diagnosis can be compromised. Such classical diagnostics techniques as MRI and CT are non-invasive and allow for faster tissue analysis, as they do not require biopsy by producing images of the body part of interest. However, even they imply several challenges. For instance, pathologists can be overwhelmed with the number of scans they have to analyse visually and, thus, be prone to less precise or even false diagnoses. In addition, the patients often have to take chemical agents to increase the contrast of the resulting scans. These contrast agents have risen concerns about their long-term safety in recent years (Wahsner et al., 2019).

The limitations that arise either directly from the diagnostic technologies or are human-related and concern the perception, the evaluation and determination of a medical condition, as well as the communication of the diagnosis, frame the following requirements to improve the clinical practice and to make the diagnostics more precise and faster:

I. Speed up the diagnostic process. It can be done if there is no requirement for the tissue specimen preparation to diagnose the presence or absence of a tumour. Alternatively, a fast and intra-operative resection and analysis of tissue are to consider. In this case, it should be possible to omit the haematoxylin-and-eosin staining or shock freezing of the tissue as these processes are time-consuming.
II. Increase the patient's safety. It can be done if low- or fully non-operative diagnostic techniques are used to avoid the tissue biopsy. Also, the sooner a tumour is detected, the better the chances to effectively treat it and to increase the patient's five-year survival rate and quality of life. Therefore, it is crucial to (further) develop early-stage diagnostic techniques. The avoidance of the so-called MRI and CT image labelling that includes the ingestion of chemical contrast agents can contribute to the increase of the patient's safety during the diagnostic process.
III. Increase the diagnostic precision and reduce the risk of insufficient and imprecise or even wrong diagnosis. This can be done by assisting the doctors with novel automated approaches during the diagnostic process. Thus, a computer-aided filter can go through a vast amount of, for instance, MRI or CT images to find the most informative ones to be evaluated by a human. Speaking of pre-processing, an automated process can adjust such image specifications as the

        contrast or correct for inhomogeneous illumination and brightness. This will produce better-quality images to be analysed by a human doctor in a further step of the diagnostic routine.

IV.    Develop tools that support decision-making. The first approach should target the automated tumour type and grade classification to assist the pathologist and radiologist. This will speed up the diagnostics process while minimizing human-related mistakes. The second approach should target the development of auxiliary tools and devices for real-time in-situ determination of tumour boundaries during surgeries. This will allow the surgeons to precisely remove the tumours while saving as much surrounding healthy tissue as possible, which will eventually lead to higher chances of patient's comparably fast and complete healing.

This list of requirements (I-IV) is by no means exhaustive. However, those are the requirements to be focused on in this overview. In the following, we will see how the application of AI to classical diagnostic methods such as, for instance, MRI or CT, targets them effectively yielding great results. We will also consider entirely novel diagnostic and imaging approaches based on various spectroscopic methods that address the requirements (I-IV). The application of AI to these methods yields much faster and less invasive diagnostic tools that will enhance tumour diagnostics and treatment in clinical practice.

## 2. Artificial intelligence in medicine

In this section, I will briefly introduce the reader to the concepts of artificial intelligence focusing on machine learning (ML) mathematical concepts. The amount of knowledge transferred in the first part of this section is sufficient to read and understand the following review of publications on AI applications for cancer diagnostics. However, the concepts of artificial neural networks (ANN), although often mentioned in the following, are beyond the scope of this review. It is enough to see them as black boxes that operate similarly to ML algorithms. If the reader is interested in deeper mathematical details and the principles of operation of ML and ANN algorithms, they consult books on these topics, suitable references are provided. In the second part of the section, I will discuss the opportunities and challenges of AI applications for cancer diagnostics. In particular, I will point the reader's attention to the problem of the scarcity and imbalance of the available cancer-related data and show some ways to circumvent it.

### 2.1 What is Artificial Intelligence?

Artificial intelligence (AI) is an interdisciplinary field of science and technology that combines topics from computer science, mathematics, robotics and engineering to develop and produce the so-called intelligent agents, i.e. artificial systems or machines that can independently, i.e. without direct interaction with the human, solve pre-defined problems. Thus, AI includes the development of autonomously driving cars, natural language processing that allows automated translation of texts and creates chatbots to interact with people, face recognition in cameras, recommendation systems on e-commerce and online streaming platforms, and automated investing in finances. Also in medicine, the deployment of AI technologies has gotten more traction over the last two decades providing the users with smart healthcare wearables, the possibility of disease evolution tracking and prediction, and improvement of the health care provision. For example, during the COVID-19 pandemics, AI-based applications helped scientists and researchers automatically extract information from a vast and exponentially growing number of scientific articles about this disease (Hutson, 2020). In particular, AI allows to predict possible and monitor already available COVID-19 cases, trace the contacts, reduce the burden on the medical staff, understand the virus entry into the body and its replication pattern by predicting its protein structure, facilitate the therapeutics and vaccine development, and curb the spread of misinformation and fake news (Arora et al., 2020). In the following, we consider the progress of AI in classical and novel spectroscopy-based applications for tumour diagnostics.

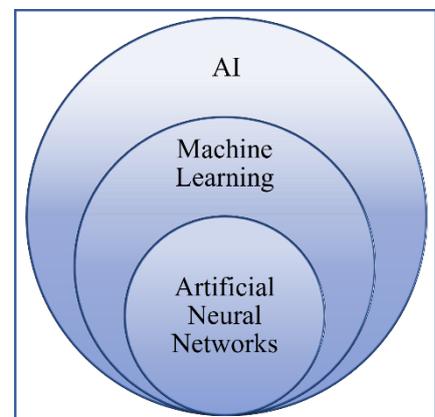

*Figure 1 Relationship between artificial intelligence (AI), machine learning (ML), and artificial neural networks (ANN)*

Machine learning (ML) and artificial neural networks (ANN) constitute subsets of AI tools and approaches [Fig. 1]. Those are computer algorithms that process quickly injected (huge) amounts of input data of sometimes various types to make predictions or decisions at the output to solve a pre-defined problem [Fig. 2]. Contrary to some software packages or plain pocket calculators, ML and ANN algorithms are not pre-programmed to make specific decisions, but derive their "knowledge" from the input data and produce an output that is data-driven and can accordingly change from input to input. Both groups of algorithms, ML and ANN, deploy concepts of linear algebra, statistics, and optimisation to unveil patterns, dependencies, and hidden structures in the input data and provide a prediction output [Fig. 2]. In particular, ANN (also referred to as *deep learning*) algorithms use a series of linear and nonlinear transformations on input data to mimic the action of firing human neurons during information processing. For more and deeper details, the reader is advised to consult such great textbooks on statistics, machine learning, and neural networks as Bishop, 2006; Du, 2014; Shalev-Shwartz, 2014; Alpaydin, 2020; Deisenroth et al, 2020.

Depending on how the ML or ANN model learning process takes place, we differentiate between *supervised* learning, *unsupervised* learning, and *reinforcement* learning. *In supervised learning*, the model learns a function to map the input data (typically a vector of the so-called *features*) to the output data (the desired vector or a scalar). The output data are labelled, i.e. equipped with the ground truth, by the so-called supervisor, usually a human with domain knowledge [1]. Having learned the dependence between the input and the output, the model is now able to predict the desired outputs using data without known labels. Supervised learning includes classification, i.e. the assignment of a categorical class value to the input data (for instance, "human" or "machine"), and regression which is the prediction of a numerical value (for example, "178" for the height or "79" for the weight of a human). Following algorithms belong to supervised learning: N-nearest neighbour (K-NN), support-vector machines (SVM), linear regression, logistic regression, ANNs in form of a multilayer perceptron, and decision trees.

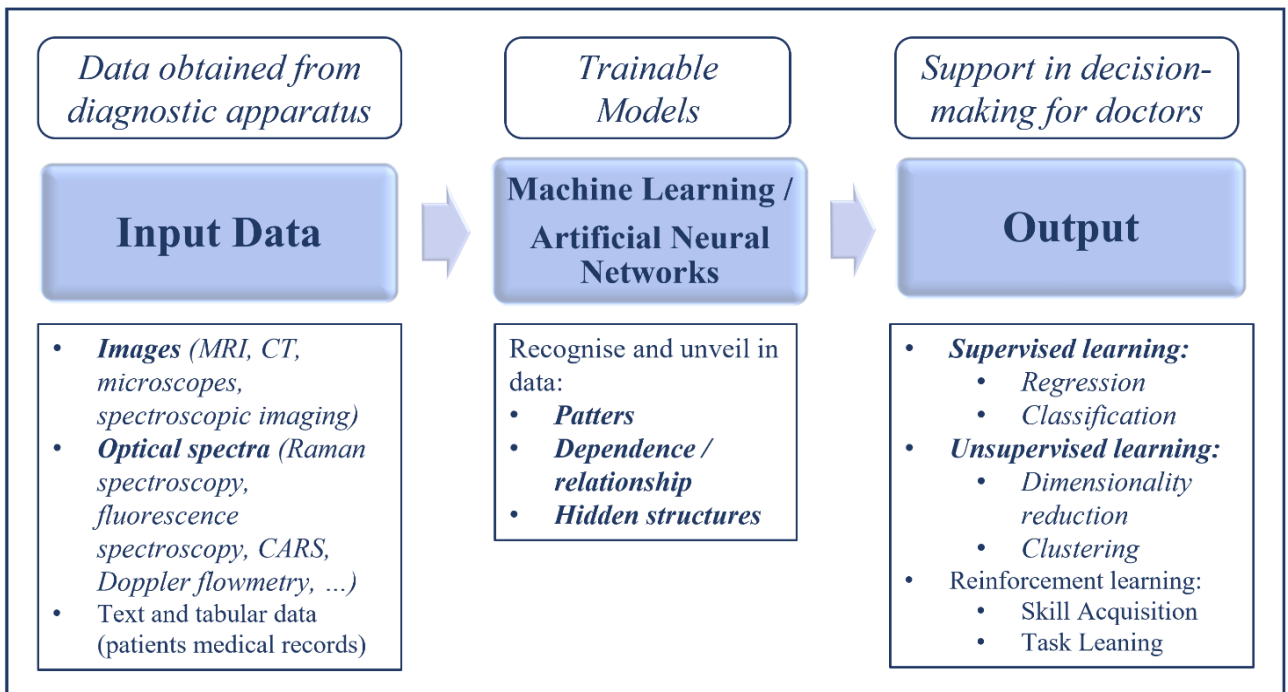

*Figure 2 Machine learning and artificial neural networks and their data throughput*

---

[1] Please, do not confuse labelling of output data for supervised machine learning with labelling in tumour imaging. In the first case, we deal with a numerical or categorical value provided by the supervisor. In the latter case, labelling denotes taking chemical agents by the patient to increase the contrast in MRI or CT images (cf. Sec. 1.2).

*In unsupervised learning*, the model learns the patterns in the data by itself without being provided with the ground-truth labels. As an example, K-means and autoencoder ANNs constitute unsupervised-learning algorithms. *In reinforcement learning*, the model learns to make a sequence of actions. The model is not taught or programmed what is the best action. Instead, it gets positive or negative rewards for its decisions. The learning aims to maximise the total reward. As an example, this type of learning can be used for autonomous driving cars to automatically determine the optimal trajectory and speed, as well as to plan motion (Du, 2014).

Please note that the diagnostic process is inherently built on the classification of patients in the class of disease-affected or healthy people depending on the availability and the level of expression of certain disease symptoms or biomarkers. Human doctors perform this classification stating the diagnosis in their clinical practice. The application of ML and ANN methods to already existing diagnostic techniques and the development of novel ones follow the pattern and, thus, mostly solve supervised classification problems. In the data-preparation stage, human doctors undertake the labelling of the output data, i.e. provide the ground-truth values [Fig. 2]. Accordingly, we will mostly see examples of supervised learning algorithms to differentiate tumour-affected samples from healthy ones. Unsupervised learning is used for data pre-processing for image segmentation tasks or feature extraction.

To develop an ML- or ANN-based intelligent agent, the programmer constructs a new data predictor, usually called a model. It can be either *discriminative* or *generative*. The discriminative model estimates a function that maps the input features $x$ to the (provided) output label $y$. Doing so, it seeks to find decision boundaries between the patterns in the data. The model training occurs by optimising the model parameters (often referred to as *weights*) to maximise the conditional probability, i.e. the probability of $y$ under the condition of $x$. These models are mainly supervised-learning ones. Examples of discriminative models are linear and logistic regression, support vector machines (SVM), classification and regression ANNs, decision trees, K nearest neighbours, and random forests.

Whereas discriminative models draw boundaries between the data groups or patterns, the generative approaches try to model how the data is placed in the data space. Thus, the generative models concentrate on the distribution of individual classes in the data and learn the joint probability of the input and the class label. Accordingly, the training of a model aims to optimise the model parameters such that the joint probability has a maximum. These are mostly unsupervised-learning approaches. Examples are linear discriminant analysis (LDA), naïve Bayes, or generative adversarial networks (GANs). Whereas the discriminative models identify tags and sort the data, the generative approaches can generate new data (Jebara, 2012; Harshvardhan et al., 2020). Both approaches are used in medical applications. The generative models are mostly used in the data preparation or pre-processing steps (especially, in the image pre-processing), whereas the discriminative models perform the diagnostics via the classification of the input data into the disease-affected and not-affected ones. In the following, we will mostly see examples of discriminative supervised-learning models. Therefore, we should look at how a *supervised model* learns.

The learning occurs in the so-called *training phase*. Here, the model uses the training set known to the programmer to self-update its parameters via a series of optimisation steps derived from the *loss function*. The general goal during this process is to minimise the model's prediction error, i.e. the difference between the values predicted by the model and the ground-truth labels. The *training dataset* is written as a set of K example-label pairs $\{(x_1, y_1), \ldots, (x_n, y_n), \ldots, (x_K, y_K)\}$ where $x_i \in \mathbb{R}^N$ are the input (feature) vectors of dimension $N$ and $y_i$ the according label vectors, $y_i \in \mathbb{R}^M$ with $M > 1$, or scalars, $M = 1$. We assume that the examples are independent and identically distributed. In this context, independent means that two data points $(x_i, y_i)$ and $(x_j, y_j)$ do not statistically depend on each other. Identically distributed means that all examples have the same probability distribution. For simplicity reasons, the input vectors are often stack to a matrix $X$, $X \in \mathbb{R}^{K \times N}$, whereas the labels are stack to a matrix $Y$, $Y \in \mathbb{R}^{K \times M}$.

Let us take an example of linear regression, a model that establishes a linear relationship between the input and output variables. In this model, the ground-truth values $y^i$ and model's output predictions $\bar{y}^i$, $i \in K$, are scalars. Their relationship reads as:
$$y^i = \bar{y}^i + \epsilon^i = w_o + w_1 x_1^i + w_2 x_2^i + \cdots + w_N x_N^i + \varepsilon^i$$
where the scalar $w_o$ is called the bias and the scalars $w_j$, $j \in N$, are called weights of the model. The bias and the weights are the trainable parameters of the model. $\varepsilon^i$ is a small deviation between the prediction $\bar{y}^i$ provided by the model and the ground-truth value $y^i$. In the matrix notation, $K$ relationships between the input $x^i$ and the label $y^i$ can be written as:
$$Y = X \cdot W + \epsilon$$
with
$$Y = \begin{bmatrix} y^1 \\ \vdots \\ y^K \end{bmatrix}, X = \begin{bmatrix} 1 & x_1^1 & \cdots & x_N^1 \\ \vdots & \vdots & \ddots & \vdots \\ 1 & x_1^K & \cdots & x_N^K \end{bmatrix}, W = \begin{bmatrix} w_o \\ w_1 \\ \vdots \\ w_N \end{bmatrix}, \text{ and } \epsilon = \begin{bmatrix} \varepsilon^1 \\ \vdots \\ \varepsilon^K \end{bmatrix}.$$

The objective is to minimise the total error $\|\epsilon\|$ that results from the contributions of the $\varepsilon^i$ components. These contributions are described by the *loss function*. In the case of the linear regression problem, the loss function can be given as a mean of the squared error:
$$L(W) := \frac{1}{K}\|\epsilon\|^2 = \frac{1}{K}\|Y - X \cdot W\|^2$$
The weights that allow the model to optimally approximate the ground-truth values $y^i$ are found via the minimisation of the loss function with respect to the weights, i.e. finding $\min_{w_j} L(W)$. This problem is often referred to as the least-squares problem. In the case of linear regression, there exists an analytical solution that provides $w_j$, $j \in K$, contributions (Deisenroth et al, 2020):
$$W = (X'X)^{-1} X'y.$$
However, for most problems, there is no analytical solution available for the loss-function minimisation. In this case, the minimisation process is achieved via an iterative optimisation scheme performed on the model weights $w_j$. The process is called *gradient descent* and it is based on the Newton-Raphson update method:
$$w_j^{new} = w_j^{old} - \eta \cdot \nabla_{w_j} L(W)$$
with $\eta$ being the learning rate, $\eta \in \mathbb{R}_+$ (Bishop 2006). During the training process, the model updates its weights such that its predictions are approximated towards the ground truth.

In the *validation phase*, the programmer trains several model versions varying model parameters. Validation serves to prove the generalisation ability of the model versions, i.e. to see how well these model versions perform on unseen data. Comparing the performance of these versions on a data *validation set*, the programmer chooses the best-performing ones. In the supervised learning approach, the validation set is built similarly to the training set and consists of example-label pairs, that were not used to train the models. The model evaluation takes place in the *test phase*. Here, the test set usually consists of examples only, whereas the labels are only known to the programmer. The outputs predicted by different model versions are evaluated by the programmer. The programmer compares the outputs of the model versions with the ground-truth labels and chooses the best performing model. This model will be used in the application (Bishop, 2006).

There is no general rule on how to split the overall dataset into the training, validation, and test subsets. Though it is important to keep in mind that the more data are injected into the model during the training process, the better it can learn. Accordingly, the biggest slice of the overall-dataset cake goes towards the training set. The rule of thumb on how to divide the overall dataset into the training, validation, and test set is $60\% - 20\% - 20\%$.

Now, let us talk about how the model performance is evaluated. In the training phase, the programmer tracks the evolution of the *training error* represented through the portion of the incorrectly predicted examples. As the loss function is minimised over several iterations, the model approximates the test data, and the training error decreases. If the training error becomes too small, we encounter the risk of *overfitting*. Overfitting means the model memorises the input data, i.e. it perfectly adjusts its parameters to the training set, which results in a small or even vanishing training error. Such a model will not be able to perform well on the input data it has not seen previously. To avoid overfitting, the programmer needs to track the *test error* that is calculated when the model is applied to the test set. While the training error constantly decreases with each iteration of the weights update, the test error will initially decrease, but at some point increase again. This point denotes the beginning of overfitting. Here, the programmer should stop the training process to balance the achieved training and test error. In general, the problem of overfitting can be mitigated by the reduction of the model complexity, introduction of regularisation parameters, and an increase of the training set (if possible). The opposite problem is the *underfitting*. In this case, the model is not sufficient to grasp the complexity of the injected data (for instance, a line is used to fit quadratic dependence in the data). An increase in model complexity by adding additional terms and parameters or an increase of training epochs (under the parallel observation of the training and test error) might mitigate the problem of underfitting (Deisenroth et al, 2020; Alpaydin, 2020).

The observation of the training and test error takes place in the training and test phase of model development. In the validation phase, the programmer has various approaches to evaluate the model performance. If the whole available is small, which is often the case in medicine or biology. The programmer can use *cross validation* to evaluate the model. The k-fold cross validation partitions the test set into *k* subsets (folds). Then, the model is trained and tested on *k-1* subsets and validated on the *k*-th one. The process is repeated *k* times permutating the subset used for validation. The average of all *k* model outputs estimates the overall model performance (Bishop, 2006; Shalev-Shwartz, 2014). Cross validation can be used for both, regression and classification.

| Metric | Formula | Meaning |
|---|---|---|
| Accuracy | $\dfrac{T_P + T_N}{T_P + F_N + F_P + T_P}$ | Classifier's overall effectiveness |
| Precision | $\dfrac{T_P}{T_P + F_P}$ | Answers the question how many examples are really positive from all examples predicted as positive |
| Recall / Sensitivity | $\dfrac{T_P}{T_P + F_N}$ | Classifier's effectiveness to identify positive labels |
| $F_\beta$-score | $\dfrac{(\beta^2+1)T_P}{(\beta^2+1)T_P + \beta^2 F_N + F_P}, \; \beta \in \mathbb{R}_+$ | Relation between truly positive labels and the ones given by the classifier |
| Specificity | $\dfrac{T_N}{F_P + T_N}$ | Classifier's effectiveness to identify negative labels |
| AUC: Area Under Curve | $\dfrac{1}{2}\left(\dfrac{T_P}{T_P + F_N} + \dfrac{T_N}{T_N + F_P}\right)$ | Classifier's ability to distinguish between the classes |

*Table 1. Summary of metrics used for validation of classification models. $T_P$: true positive, $T_N$: true negative, $F_P$: false positive, $F_N$: false negative (Sokolova & Lapalme, 2009)*

For classification specifically, we can use one or several metrics summarised in Tab. 1. Those are accuracy, precision, recall (sensitivity), $F_\beta$-score, specificity, and AUC. The closer a classifier approaches the value of 1 or 100% in each metric, the better its performance.

In the following, we will use the term AI interchangeably for both, only the ML/ANN data analysis techniques and bigger implementations of diagnostic techniques and tools that take advantage of ML and ANN to target the requirements I-IV (Sec. 1.2). Also, the reader will encounter names of several different ML or ANN algorithms used in cancer diagnostics. If the reader is interested in the mathematical framework of these algorithms, they can consult various open online resources of their choice. I suggest visiting a free library called scikit-learn (*scikit-learn.org*) for ML algorithms and examples of their usage and TensorFlow (*tensorflow.org*) for ANN definitions and applications as these platforms are not only exhaustive but also trustworthy.

## 2.2 Artificial Intelligence in Medicine: Possibilities and Challenges

As in other fields of technology, AI is now quickly invading medical diagnostics as it enables automated processing of vast amounts of medical data, supports the doctors in their decision-making and provides opportunities for real-time in- and ex-vivo diagnostic solutions. In doing so, AI-based approaches target the challenges of cancer diagnostics we discussed in Sec. 1.2. As we will see in this section, they show the potential to increase the speed of the overall medical process while minimising risks for the patient. The recent development of AI-based techniques in medicine is enhanced by the increasing availability of various resources. Thus, computers are becoming faster while getting cheaper. Online teaching platforms such as, for instance, Coursera (*coursera.org*), Udemy (*udemy.com*), or EdX (*edx.org*) allow professionals from different fields, including medical personnel, to gain knowledge about machine learning and neural network methods and acquire skills in statistics and programming languages (Abedi & Beikverdi, 2012; Zinovieva et al., 2021). Various free machine-learning software libraries such as, for instance, Scikit-learn (*scikit-learn.org*) or TensorFlow (*tensorflow.org*) facilitate a quick start in solving automated data classification, regression, or clustering tasks. This means that you do not need to be a computer scientist anymore to be an active citizen of the AI-driven world. You can be a doctor or, for instance, a scientist in biology or medicine and yet be able to use AI-based techniques on a professional level. All these aspects have resulted in the growing number of scientific articles on AI for medical applications in recent years some of which I will review in the further course concentrating on the field of tumour diagnostics (Haleem et al., 2019; Castiglioni et al., 2021; Ramkumar et al., 2021).

So far, the biggest part of AI problems in medical diagnostics has accounted for supervised learning with the target to classify tissue samples or body part images into normal and affected by a disease. In cancer diagnostics, the doctors want to classify tissue samples into healthy and cancerous ones. Also, they would like to have segmentation of a diagnostic image into parts that present the tumour and other kinds of tissue. To target these types of problems, we need tags or labels that give us the ground truth about the tissue type and quality. This information is provided by human professionals such as surgeons or pathologists as they carry factual knowledge about the tumours. Doctors mostly solve classification problems in their daily practice when they determine the diagnosis using, for instance, MRI or CT images. They, however, use their expertise, experience, and knowledge to diagnose. AI can support them in their classification and decision-making tasks through the automation and speed-up of this process.

Although AI is on the rise in medicine, there is still a problem with the availability of suitable data to train machine learning or ANN models. Usually, biological data are scarce due to a low number of available samples from which the data can be collected. The possibility to collect and use patients-related data relies on patients' consent and is liable to regulations concerning the patients' safety (Del Carmen & Joffe, 2005; Yip et al., 2016; Artal & Rubenfeld, 2017) and the personal data protection rules such as, for instance, the General Data Protection Regulation (GDPR) framework in Europe (Rumbold & Pierscionek, 2017; Voigt & Von dem Bussche, 2017). Accordingly, the size of the datasets to be used for training and testing ML and ANN models is often limited which can result in the so-called *overfitting* of the AI models and the subsequent lack of generalisation on a new dataset. Apart from trying to increase the dataset size by adding more data, various so-called data augmentation techniques can be applied to mitigate the problem of limited data. In image processing, image translation, rotation, shifting, magnification as well as changes in the brightness can effectively augment the dataset (Pradhan et al., 2020). In time series such as, for instance, EEG brainwaves traces, the augmentation can be done via the generation of a fourth artificial sample using three real ones using the following analogy: the fourth (artificial) sample relates to the third one as the second to the first (Lotte, 2015). Adding some additional noise can help to augment time-series or spectral data. Using models that are pre-trained on large benchmark datasets such as, for instance, Inception, VGG, and U-Net (Ferreira et al., 2018; Kang et al., 2021; Saxena et al., 2021) can mitigate the problem of small image datasets specifically in the case of deep learning and ANNs (Pradhan et al., 2020).

The second problem that directly connects to the limited dataset volumes lies in the imbalance of data in a dataset meaning an unequal distribution of data samples across the classes. In this case, the AI models are affected towards the majority class during the training and falsely predict the examples from the minority classes in the test and application phase. There are several approaches to mitigate the problem of unbalanced datasets. Thus, on the data level, one can increase (over-sample) the number of examples in the minority classes just by copying the minority-class examples back in the dataset. As opposed to this approach, it is possible to under-sample the majority class by deleting some data examples in it. In addition, it is sometimes possible to use computer simulations to generate additional artificial samples to enrich the minority class and enlarge the whole dataset. However, high reliance on artificially generated data can limit the generalisation of the AI models when it comes to real data. Apart from the attempt to achieve a more balanced distribution on the data level, it is also possible to tweak the model level. Thus, one can introduce penalisation of the loss function by weights that are defined in dependence on the number of samples in each class. This approach, however, requires the possibility to define the loss function, which is not always possible. Many researchers report the combination of data-level and model-level approaches to target the problem of an imbalanced set (Pradhan et al., 2020).

In the next section, we will take a look at the application of AI in tumour diagnostics. We will see that, despite various and numerous challenges, the application of ML and ANN techniques to classical diagnostic methods such as, for instance, MRI or CT shows highly promising results in the support of human doctors in their decision-making during the early-stage or pre-operative diagnostics. The development of spectroscopic techniques that include implementations of ML or ANN approaches allows for fast and low-invasive intra- and post-operative tissue analysis. Although the spectroscopy-based technology is not mature enough to be deployed routinely in clinics as of 2022, there is hope that it is only a matter of a few years for it to evolve into high-performing and fast implementation for everyday clinical practice.

## 3. Artificial intelligence for classical tumour diagnostics methods

Here, we will consider well established imaging methods such as, for instance, MRI and CT, and see examples of how the AI progresses in this field. MRI (magnetic resonance imaging) is one of the most commonly and broadly used non-invasive diagnostic methods for cancer detection. It utilises strong magnetic fields and their gradients to induce spin polarisation and the resulting radiation of radio frequencies in hydrogen atoms to produce images of inner organs (Katti et al., 2011). There are two types of images generated by MRI, the so-called T1-weighed and T2-weighted images. T1 and T2 are the relaxation time periods between the magnetic pulse excitation and the taking of the images. The T1-weighted scans are produced using short repetition and echo times, whereas T2 scans deploy longer repetition and echo times. The choice between T1 and T2 depends on the tissue properties to be analysed (Katti et al., 2011; Feng et al., 2013; Fischbach-Boulanger et al., 2018; Cooley, 2021). A further method includes diffusion-weighted imaging that deploys the diffusion of water molecules to generate contrast in MRI scans (Bammer, 2003). Dynamic contrast-enhanced MRI analyses the temporal enhancement pattern of tissue following the introduction of a paramagnetic contrast agent into the vascular system. For this, baseline images without contrast enhancement are taken first. Then, they are followed by a series of images acquired during and after the arrival of the contrast agent (Gordon et al., 2014). In recent years, different MRI methods are combined in the so-called multiparametric MRI (mpMRI) to achieve better diagnostic results. Thus, instead of deploying one of the methods, mpMRI approach can include, for instance, T2-weighted, diffusion-weighted, dynamic contrast-enhanced and even MR spectroscopy images (Demirel & Davis, 2018). As MRI generates images, artificial neural networks (ANNs) are the natural choice to analyse them, although not the exclusive one. Therefore, the number of published articles that mention the application of neural networks for MRI scan processing has widely outperformed the number of articles that report other AI approaches and is constantly growing since 2015 (Hajjo et al., 2021).

Currently, the MRI-based diagnostic power of AI is still often under the performance of experienced radiologists and oncologists. However, the quickly evolving approaches (sometimes deploying pre-trained neural networks used for image processing such as, for instance, VGG16 and VGG19, U- or V-net, GoogleNet (Hu et al., 2020; Lundervold & Lundervold, 2019; Pradhan et al., 2020) are highly promising to evolve to the level of a human professional or even to surpass human expertise. This will allow for the automation of simple tasks such as image classification and assist the radiologist in the decision-making in complicated cases.

Let us now take a look at some examples of recent advances in the field of AI application to MR imaging. Tang et al. report the application of a random forest model and a two-layer neural network to T1-weighted brain scans with 512 grey levels to classify different brain tumour types and predict tumour growth rate in a preclinical mouse model. The mouse model is compared with human tissue. They achieved 92%, 91%, and 92% in specificity, and 89%, 85%, and 73% in sensitivity when classifying the tumour types in mouse glioma, human glioma, and human medulloblastoma. The overall accuracy was 84%. Using an ANN, the authors were able to predict tumour growth with a mean square error of 16%. The authors are convinced that the addition of more data would improve the performance of their models, especially taking into account that both algorithms, random forest and the considered neural network, heavily rely on the amount of the training data (Tang et al., 2019).

Zhen et al. write about the development of a deep learning system for the diagnosis of liver tumours by first classifying liver focal lesions into seven categories such as cyst, haemangioma, focal liver lesion, other benign nodules, hepatocellular carcinoma, metastatic malignant tumours from other sites (colorectal, breast, lung, pancreas, etc.), and primary hepatic malignancy other than hepatocellular carcinoma. Subsequently, the authors developed a convolutional neural network (CNN) to classify benign and malignant tumours using unenhanced MRI sequences. The CNN models performed well in distinguishing between malignant and benign liver tumours (AUC = 0.946). The CNN that used unenhanced images combined with clinical data greatly improved the performance of classifying malignancies as hepatocellular carcinoma (AUC = 0.985), metastatic tumours (AUC = 0.998), and other primary malignancies (AUC = 0.963). To ensure good training of the CNN models, the authors used data collected from 1210 liver-tumour patients. Zhen et al. consider their results as proof of the feasibility and potential superiority of deep-learning systems in liver tumour MRI-based diagnostics. However, they acknowledge that their studies did not include the whole spectrum of possible liver lesions and excluded less common liver masses such as abscesses, adenomas, and rare malignancies (Zhen et al., 2020).

Hu et al. report the fusion of information obtained from sequences of dynamic contrast-enhanced and T2-weighted MRI scans for the detection of breast cancer. The fusion of information occurred at three different levels, namely as image fusion, feature fusion, and classifier fusion. The authors pre-trained a CNN to extract features from the MRI scans. An SVM classifier subsequently distinguished between benign and malignant lesions using the extracted features. Using AUC as a metric, the authors achieved the following values for each information fusion method: AUC (image fusion) = 0.85, AUC (feature fusion) = 0.87, and AUC (classifier fusion) = 0.86. These results outperformed the classification values achieved using only one type of MRI scans, either dynamic contrast-enhanced or T2-weighted images. Accordingly, it is to conclude that the combination of several modalities to fuse the information can increase the performance of AI models (Hu et al., 2020).

The review by Bardis et al. lists the results of the application of CNNs to mpMRI images for the detection of prostate cancer. Thus, they reported AUC values between 0.90 and 0.97 depending on the mpMRI modalities, i.e. on what MRI methods were taken into account (Bardis et al., 2020). The authors point out that, in the case of prostate lesion detection, satellite small lesions can be challenging to detect. However, they are optimistic that the deployment of evolving AI techniques will assist the radiologist in the decision-making, although never replace them fully (Bardis et al., 2020; Tătaru et al., 2021).

Also in the case of lung cancer detection using computer tomography (CT) imaging, we see highly promising results of the AI application. For this type of cancer, 70% of all cases are detected at the later stage when the tumours are already metastatic and the treatment is difficult. This occurs because people often neglect the first symptoms of lung cancer such as cough and fatigue. The mortality rate is accordingly high making lung cancer the deadliest among all cancer types. Thus, 75% of people diagnosed with lung cancer die within 5 years after the diagnosis. CT is one of the gold-standard methods to diagnose lung cancer (Mazzone et al., 2021). It uses radiation from an X-ray source to generate cross-sectional (tomographic) images of a body and allows physicians to detect even small lesions in the lungs (Doria-Rose & Szabo, 2010). Regular yearly CT scans of people at risk such as current and former smokers would allow discovery of the tumours at an early stage and, thus, increase the 5-year survival rate by 20-30%. At the same time, the amount of information to analyse might overwhelm the capacities of the hospitals (Svoboda, 2020). AI is a promising tool to automatise scan evaluation and support radiologists in their diagnostic routines. Thus, Sun et al. report an accuracy level of 0.8119 in a deep belief neural network that was designed to diagnose lung cancer from CT images (Sun et al., 2016). Chamberlin et al. developed two CNN modules one of which identifies lung nodules with an excellent agreement with radiologists. Thus, their AI network achieved a Cohen's kappa coefficient of 0.846 which denotes an almost perfect agreement with real doctors, as well as a sensitivity value of 1 and a specificity value of 0.708 (Chamberlin et al., 2021). To analyse CT scans for lung cancer, not only neural networks can be used, but also other machine learning techniques. Bi et al. review a list of articles that deploy support-vector machines (SVM) and random forests to predict lung cancer in screenings. The authors mention a random forest with AUC = 0.83 and an SVM with an accuracy of 73% for this task among other interesting results on the differentiation between benign and malignant lung lesions and the prediction of the mutational status (Bi et al., 2019).

Breast cancer is one of the most widely spread cancer types among women. Thus, one in eight American women will be diagnosed with breast cancer at some time during their life according to the National Cancer Institute (Howlader et al., 2020). Early diagnostics is therefore decisive to prevent worst-case scenarios and increase the success of medical treatment. A well-established non-invasive technique for breast cancer detection is mammography. It uses low-energy X-rays to produce images of the breast tissue (Gøtzsche & Jørgensen, 2013). The application of convolutional ANNs to mammographic scans allows to automatically differentiate between normal and abnormal tissue with an accuracy value of 94.92%. Also, ANNs can classify benign and malign mammograms with an accuracy of 95.24% achieving a sensitivity value of 96.11% (Ayer et al., 2010; Zhang et al., 2020). Infrared thermal imaging is a further non-invasive method that has a high potential to speed up the breast-cancer diagnostic process when combined with AI (Hakim & Awale, 2020; Zadeh et al., 2020). As a breast with cancer always exhibits an increase in thermal activity in the precancerous tissues and areas that surround the developing breast cancer, the analysis of thermal scans by various AI techniques suggests a high level in the healthy vs. cancerous tissue differentiation (Mambou et al., 2018). Indeed, Vardasca et al. report classification accuracy values going up to 98% for various AI techniques (ANN, SVM, decision trees, AdaBoost, k-NN, etc.) when applied to thermal images for breast cancer diagnostics (Vardasca et al., 2019).

In this section, we reviewed and discussed how the application of AI can assist cancer diagnostics. We conclude that the machine still mostly performs worse in the detection of cancer or cancer grade from MIR or CT images or spectroscopic analysis of biopsy material than experienced pathologists and oncologists. However, it is only a matter of time before it starts outperforming human assessment in straightforward classification tasks and in the automated processing of vast amounts of data helping doctors provide faster and more precise diagnostics. Advance in the modelling methods and obtaining more data with more precise measuring algorithms to further develop medical signal databases is an inevitable factor in future progress (Goldberger et al., 2000; Trifirò et al., 2018; Horton et al., 2019).

However, it is not only the problem of data acquisition and quality that slows down the advancement in cancer diagnostics. It is the technology itself. The apparatus of widely used MRI or CT is bulky and expensive to build and maintain. Also, the personnel has to be protected from strong magnetic fields of MRI or X-rays of CT. This makes these technologies difficult to be used intra-operatively, i.e. during surgeries, as the patients have to be transferred to the scanners whereas the personnel are removed during the process of imaging. Additionally, electric noise that is radiated from anaesthesia and patient monitoring systems can negatively influence the quality of imaging and, thus, compromise the tracking and assessment of a patient's vital parameters (Gandhe & Bhave, 2018). Some ground-breaking progress occurred in the last years as mobile MRI and CT scanners were introduced to support brain and spinal surgeries (Ashraf et at., 2020; Liu, 2021). When compared to their stationary counterparts, these scanners are smaller, less expensive, and their radiation is lower and, thus, less dangerous for the doctors and operating personnel. However, the image quality is inferior and needs AI correction (Gandhe & Bhave, 2018). Also, the field of their application is still reduced to the brain and spinal surgeries as I already mentioned. So, we still need imaging and tissue-information collection techniques that can be fast, low-invasive and safe, low-cost, and allow for intra-operative diagnostics and doctors' support. As we will see in the next section, spectroscopy-based techniques aim at these challenges and have already shown promising results in combination with AI.

## 4. Novel tumour diagnostic methods: spectroscopy is on the rise
### 4.1 Review of *ex-vivo* spectroscopic approaches for tumour diagnostic

So far, we considered various non-invasive tumour imaging technologies that, in combination with AI methods, constitute promising diagnostics techniques in early-stage, pre-operative, and with some restrictions on intra-operatively diagnostics. Now, we consider optical, spectroscopy-based approaches that are on the rise to significantly speed up the diagnostic process. The principle of spectroscopy lies in the illumination of a material with the light of a certain wavelength and the subsequent collection of reflected or re-immitted optical spectra. These spectra contain information about the material's physical and chemical qualities. Spectroscopic methods allow for imaging and chemical composition analysis of biological tissue and can be, thus, used for diagnostic purposes. Contrary to classical methods such as MRI or CT, the data collection is fast. Namely, it takes only a few minutes to collect optical spectra from the tissue under study. Subsequent application of AI methods allows for fast evaluation of spectra. In terms of time consumption, the spectroscopy-based diagnostic methods promise to outperform classical approaches in tumour diagnostics. Spectroscopy of biological tissue does not require the administration of chemical agents to enhance the imaging quality. As we will see in the further course, spectroscopic probe implementations allow for low- and even non-invasive data collection. These aspects contribute to the minimisation of patients' health risks. Last but not least, spectroscopy-based approaches are compact compared to CT or MRI scanners and can be deployed directly during surgeries without danger for the patient and the operating personnel. In the following, we will see examples of how spectroscopy contributes to the diagnostic analysis of body fluids, enhances tissue imaging, and allows for fast differentiation between healthy and cancerous material.

### 4.1.1 Mass spectroscopy for body fluids analysis

In oncology, the extraction of blood or biopsy material, as well as the intra- or post-operative tumour tissue analysis, are often required to understand the tumour development and to apply patient-specific treatment to target the disease. Body fluids such as blood or/and plasma are simple to extract and are extremely informative for cancer diagnostics and development tracking as they might contain cancer-type-specific biomarkers and freely circulating cancer cells. As in the case with other tumour diagnostic techniques, AI has invaded the realm of blood analysis delivering promising results in the detection and prediction of certain types of cancer. Let us now take a look at some examples.

Mass spectrometry is getting traction as it provides rapid and precise measurements or relative abundance of certain proteins in a specimen. It generates particle mass spectra as intensity vs. mass-to-charge ratio of sample ions (McDonnell & Heeren, 2007; Xiao et al., 2020). In cancer diagnostics, mass spectra of blood samples encode information about the range of protein biomarkers that can be related to a tumour. The spectra are accordingly complex and extremely fragile to treat in order not to lose important biomarker imprints. Already in the 2000s, such AI techniques as ANNS, SVM, KNN, decision trees, LDA, Naive Bayes and various ensemble methods have shown their ability to decode the information stored in the mass spectra and to classify the spectra coming from healthy people and people who are tumour-affected (Shin & Markey, 2006). Apart from the cancer biomarkers, freely circulating tumour cells (CTCs) in blood have a high level of prognostic importance in cancer diagnostics and treatment. However, it is problematic to determine the origin of the CTCs as they can originate from both, the primary and metastatic lesions. Apart from that, they are extremely rare. Normally, cancer patients usually exhibit 10 CTCs per 1 ml of blood. Nissim et al. report the deployment of interferometric phase microscopy, a holographic microscopy method that provides morphological and refractive imaging of cells, to produce images of colorectal CTCs in a flow of blood (Barnea et al., 2018; Nissim et al., 2021). The images were then processed and classified taking into account their morphology and quantitative phase features during the cell flow. Using SVMs, the authors achieved an accuracy level of 92.56% in the differentiation between four types of blood cells and two types (primary and metastatic) of colorectal CTCs (Nissim et al., 2021). This is an extremely promising result considering the rareness of the CTCs and the related problem of sufficient data collection. Both techniques, mass spectrometry and interferometric phase microscopy are low invasive and can be applied in real-time as they only require a small amount of patient blood. They are also label-free which means the patient does not need to take any chemicals to enhance the procedure. These aspects might speed up the diagnostic process while reducing the risks for the patient.

### 4.1.2 Optical spectroscopy and AI-assisted tissue preparation

Often, the extraction and histopathological analysis of a biopsy are required to increase the level of diagnostic accuracy. For instance, in the case of gastric cancer, histopathological analysis allows for accurate detection of lymph node metastasis (Matsumoto et al., 2019). Digital histopathology allows for scanning and further storage of biopsy specimen information in form of images. Here, AI comes into question for automated classification between different types of biopsy tissue. Thus, Arunachalam et al. analyse microscopic images of haematoxylin-and-eosin stained tissue slides by 13 AI algorithms to detect osteosarcoma and to differentiate between the regions of the non-tumour, necrotic tumour, and viable tumour. The authors achieved accuracy levels of over 90% for each tissue-type class. The scans they used were annotated by experienced pathologists to provide sample labelling for supervised training of the AI models. The authors stress that the accurate detection of specifically tumour necrosis using AI will significantly improve the success of osteosarcoma treatment as tumour death in response to pre-operative chemotherapy has been a significant prognostic indicator for several decades (Arunachalam et al., 2019).

As pointed out in Sec. 1.2, the gold standard of sectioning and haematoxylin-and-eosin staining the biopsy tissue to enhance its cellular and structural details is a time-consuming process and can take at least 30 minutes to be performed. This is an obstacle if more knowledge is required directly during surgery, i.e. intra-operatively. Optical spectroscopy combined with AI comes into question to significantly speed up the process of tissue analysis by dropping the haematoxylin-and-eosin staining step. Here, tumour biopsy samples are exposed to light at certain wavelengths (depending on the spectroscopy method) and the optical response of these samples is collected in form of optical intensity spectra. The spectra are subsequently analysed to retrieve the bio-chemical and morphological information about the tissue under study.

Matsumoto et al. suggest omitting even the sectioning, i.e. thin-slicing, of the resected biopsy samples by deploying deep-UV excitation fluorescence microscopy. Fluorescence is the process of light re-emission by the material that is excited by light of a certain wavelength or in a certain spectral range. The fluorescent spectra are often red-shifted with respect to the excitation central wavelength as the exciting light loses some of its energy in the material under study due to the relaxation process in the vibrational energetic states. As the vibrational states are material-specific, the fluorescence spectra carry information about the molecular composition of the tissue (Kim et al., 2020). In the authors' approach, the tissue sectioning becomes obsolete, as the volume of the material excited via deep-UV is limited to a thin layer near the tissue surface. This happens due to the narrow depth of the light penetration in this spectral range. As the tissue fluorescence spectra excited by deep-UV allow for well-resolved imaging of the cell nucleoplasm, nucleolus, and cytoplasm among other tissue properties, also the haematoxylin-and-eosin staining to visualize the cells and the tissue morphology can be dropped, not only the thin-slicing. This will significantly speed up the sample preparation step in the diagnostic process. The authors achieve a further increase in the processing speed by deploying pre-trained ANN models such as VGG16, Inception v3, and ResNet v2 for the detection of metastasis in deep-UV microscopic images. The images were obtained from lymph node biopsies in patients with gastric cancer. The authors achieve accuracy values of 98.2 – 98.8% in the differentiation between metastasis-positive and metastasis-negative lymph-nodes images. Having these highly promising results, they are convinced, that deep-UV fluorescence excitation in combination with AI is a suitable technique for intra-operative rapid diagnostics (Matsumoto et al., 2019).

Bocklitz et al. propose to produce pseudo-haematoxylin-eosin-stained images of colon cancer in mice by a combination of several nonlinear spectroscopic modalities to generate the images of the tissue samples and subsequent AI-based processing. Thus, the images were taken in different spectral ranges using such nonlinear imaging techniques as the Coherent anti-Stokes Raman scattering spectroscopy (CARS), the two-photon excited fluorescence (TPEF) spectroscopy, and the second-harmonic generation (SHG) spectroscopy. CARS resembles Raman spectroscopy in targeting vibrational modes in molecules but uses several photons to excite them by deploying simultaneous radiation and focusing of three laser beams at different wavelengths (Tolles et al., 1977; Roy et al., 2010). The authors adjust this spectroscopic modality to map the CH2 distribution to visualise the lipids in the tissue. TPEF uses two low-energetic photons to excite a biological tissue to produce fluorescence spectra in a higher-energy spectral range (Helmchen & Winfried, 2005; Zheng et al., 2011). The authors collected the TPEF signal in the spectral range of 426–490 nm coming from such strong autofluorophores as elastin, NAD(P)H, and keratin. SHG is a process of doubling the frequency of light when it passes through certain nonlinear media (Schneider & Peukert, 2007). Here, the second harmonic was generated by the fibrous collagen network in the tissue. The authors collected its light at 415 nm. Overlapping the images from CARS, TPEF, and SHG, Bocklitz et al. produced high-contrast detailed images of resected colon specimens in false colours. Then, using a partial least-squares regression (PLS) model, they generated pseudo-haematoxylin-eosin-stained images that look like images from real haematoxylin-and-eosin stained tissue samples although the step of the staining itself was fully omitted. The generation of pseudo-haematoxylin-eosin-stained images by several nonlinear spectroscopic modalities and the subsequent AI-based processing takes only a few minutes as opposed to the process of real tissue staining. The pseudo-haematoxylin-eosin-stained images are of such high quality that they can be easily analysed by human pathologists or various AI methods in a further step (Bocklitz et al., 2016).

**4.1.3  Optical spectroscopy and AI help differentiate between cancerous and healthy tissue**

The removal of a tumour is a difficult and delicate process. Thus, the surgeons try to fully remove the tumour while saving as much healthy material as possible. However, their view is often limited by body fluids such as blood or gall so the surgeons often need to rely on their expertise and intuition to determine the exact boundary of the tumours. A spectroscopic and AI-based approach that allows real-time point-by-point differentiation between the healthy and cancerous tissue would yield more precise and probably faster tumour removal and, thus, significantly increase the chance of a full recovery or at least enhance the 5-years survival rate and patient's quality of life after the surgery.

Recently, there have been several spectroscopy-based approaches introduced to help surgeons classify the material intra-operatively in near future. For now, most of these approaches are being developed and tested *ex-vivo*, i.e. using resected tissue. Thus, Bogomolov et al. pursue the development of an intra-operative spectroscopic method for kidney cancer diagnostics by analysing frozen specimens of healthy and cancerous renal tissue samples combining fluorescence and mid-infrared (MIR) fibre spectroscopy. Whereas fluorescent spectra contain information about some known cancer biomarkers, spectroscopy with mid-IR light allows detecting various organic molecules via the excitation of the fundamental vibration frequencies of their functional groups. The authors excited tissue fluorescence with a 25-mW laser at the wavelength of 473 nm, whereas the mid-IR spectra were collected with a Matrix MF spectrometer. Analysing the fluorescent spectra, the authors identified flavin adenine dinucleotide (FAD), collagen, and porphyrins as the most pronounced biomarkers in the kidney tissue samples under study. MIR delivered complementary chemical information. Applying PCA and partial least-squares discriminant analysis (PLS-DA) to the fluorescent and MIR spectra, the authors achieved accuracy values of 98 – 99% in the classification of healthy and cancerous renal tissue samples (Bogomolov et al., 2017).

A further approach that was recently studied to facilitate intra-operative analysis of the tumour tissue samples is based on Raman spectroscopy. Raman spectroscopy is a powerful technique to produce optical spectra that encode information about the molecular vibrational modes allowing to precisely identify the molecules involved (Cordero et al., 2018). Assuming that tumours have different protein abundance than the surrounding healthy tissue, one can expect to see a difference in Raman spectra collected from cancerous and healthy tissue (Austin et al., 2016; Jermyn et al., 2016). Raman spectra from biological material are, however, complex and difficult to be decoded by the eye. Here, AI techniques assist the doctors in quickly obtaining information about the tissue under investigation. Thus, Talari et al. report the usage of Raman spectroscopy in combination with principal component and linear discrimination analysis (LDA) to detect breast cancer. The tissue samples were prepared and structured into tissue microarrays (TMAs) and excited at the wavelength of 532 nm to obtain Raman spectra. LDA was able to classify the samples into the breast cancer subtypes such as luminal A, luminal B, HER2, and triple-negative with predicted specificity accuracy of 70%, 100%, 90%, and 96.7%, respectively (Talari et al., 2019).

Tissue microarrays (TMAs) are quickly taking over the field of biopsy preparation, but they are time-consuming to be produced and hardly possible to be used intra-operatively as they usually constitute arrays of formalin-fixed paraffin-embedded or frozen samples (Jawhar, 2009). Contrary to Talari and their co-authors, He et al. work towards a direct intra-operative deployment of Raman spectroscopy by using untreated biopsy samples and omitting the step of the TMA sample preparation. Thus, the authors collect optical Raman spectra from kidney tissue at the excitation wavelength of 785 nm. For this, the surgeons took normal kidney tissue, fat, and tumour tissue samples with a diameter of 2 mm directly during the partial or complete kidney removal. Every sample was divided into two parts, one for the acquisition of Raman spectra and the other one for pathology analysis to label the spectra. Raman spectra and pathology results were used to train an SVM classification model. The developed model was able to distinguish between the kidney tumour from the fat and normal renal tissue with a test accuracy of almost 93% and to classify the tumour subtypes and tumour grades with a test accuracy of 86.79% and 89.53%, respectively (He et al., 2021). Kothari et al. use two ML algorithms, first k-means and subsequently a stochastic nonlinear ANN, for autonomous classification of Raman spectra obtained from resected female breast tissue. In this scheme, k-means clusters cancerous and non-cancerous spectra with maximal accuracy of 94.6%. The neural network generates probabilities of the correctness of the clustering (Kothari et al., 2011). Raman spectroscopy is a powerful technique as it allows to study of the molecular composition of the tissue. However, Raman spectra from biological samples are complex and suffer from a low signal-to-noise ratio. He et al. introduce a variational autoencoder (a certain type of ANNs) to downscale the complexity of Raman spectra and reduce their noise. After that, the features obtained by down-scaling the Raman spectra are fed into different algorithms to find the best-performing in the classification of kidney samples extracted from human patients with kidney cancer. The authors found that a Gaussian naïve Bayes model is best performing with an accuracy of 81.4% (He et al., 2022). Due to its simplicity in operation, but richness in the information load, Raman spectroscopy will certainly gain more attention and applicability in medicine in the next years. Therefore, it is important to find ways to treat and reduce the complexity of Raman spectra. Although the final cancer classification results achieved by He et al. are inferior to what we have already seen. Their approach for Raman spectra reduction in form of a variational autoencoder gives rise to a broader discussion in the spectroscopic data pre-processing and certainly deserves attention.

Hyperspectral imaging (HSI) is a highly promising technique for fast intra-operative determination of tumour boundaries. It allows examining material structure by collecting light over a broad spectral range (Lu & Fei, 2014). Various AI technologies show their potential in a wide range of research fields (Ozdemir & Polat, 2020). Several research groups have reported promising results for oncology. Using hyperspectral images obtained with a Xenon white-light illumination source from resected tissue samples and analysed with a convolutional ANN, Halicek et al. were able to differentiate between head and neck cancer material and healthy tissue with an accuracy value going up to 96.4% (Halicek et al., 2017). Beaulieu et al. report the collection of spectra of freshly resected colon specimens using a visible-NIR spectrometer with 1 nm resolution for wavelengths between 350 nm and 1800 nm. The collected fluorescence spectra underwent LDA analysis for classification between healthy tissue and colorectal cancer. The authors achieved values of 61.68% in sensitivity and 90% in specificity in the case when the spectra were collected from outside cancer (extraluminal collection) and even sensitivity of 91.97% and specificity of 90% for within (intraluminal) collection (Beaulieu et al., 2018).

Spectral imaging does not always need to be hyper-spectral to be efficient for cancer diagnostics. In addition, only a few monochromatic images taken at different wavelengths can suffice to provide enough information for the classification of cancerous and healthy tissue. This is because lipids, proteins, and fat undergo optical excitation in different spectral ranges. Thus, Bondarenko et al. show that three monochromatic images taken in the spectral range between 450 nm and 950 nm and classified with a convolutional ANN provide us with a highly efficient way of skin melanoma diagnostics. Thus, the authors achieved an F1 score of 93% with the ANN they developed to classify the monochromatic images (Bondarenko et al., 2020). Using the same spectroscopic approach as well as a convolutional ANN, this group showed that this combination can be successfully deployed to differentiate between head and neck cancer tissue and other non-neoplastic tissue types (Rodner et al., 2018).

The examples of spectroscopic approaches presented in this section show high potential to enhance diagnostic speed and precision. However, they all were tested ex-vivo, i.e. using extracted blood or resected tissue. In the following, we will see successful implementations of spectroscopy-based tools tested in-vivo, i.e. in human cancer patients.

### 4.2 In-vivo Approaches for Cancer Detection

Previously, we discussed various approaches to how the application of AI assists cancer diagnostics. Some of them accompany already well-established imaging diagnostics techniques such as mammography, MIR, or CT. Novel approaches use spectroscopic methods to collect optical spectra from the tissue under study and to evaluate these spectra by means of machine learning or neural network algorithms. As we have seen, recently published spectroscopic methods used either resected, well-prepared or conserved tissue samples or extracted and analysed the tissue intra-operatively, i.e. *ex-vivo*. In the best-case scenario, the *ex-vivo* tissue analysis can take minutes. In the worst-case scenario, it can be hours or even days, especially if the sample slicing and staining are involved. Let us now consider approaches that are developed for *in-vivo* diagnostics. The first group of approaches includes spectroscopy-based tools that are implemented as endoscopic systems and can be used non-operatively via body orifices. The second group includes fibre probes and needles that can be used by the surgeons intra-operatively for the determination of tumour material and boundaries of inner organs.

A spectroscopy-based endoscopic system deploys an excitation (diode) laser and optical fibres to guide the laser light to the tissue and collect its spectra for analysis. As we have seen previously, Raman spectroscopy is a powerful technique that detects the molecular composition of the tissue and, thus, allows for differentiation between cancerous and healthy material. The authors Žuvela et al. and Shu et al. report a fibre-optic endoscopic system that deploys Raman spectroscopy to obtain spectra excited at 785 nm. They test this system *in-vivo* in patients with nasopharyngeal carcinoma and healthy people. Collected Raman spectra are evaluated with a partial least-squares LDA model achieving an accuracy of 98.23%, sensitivity of 93.33%, and specificity of 100% for nasopharyngeal carcinoma classification (Žuvela et al., 2019). In one of the follow-up papers, the authors develop a CNN to identify nasopharyngeal carcinoma in Raman spectra obtained from three groups of people: patients affected by nasopharyngeal carcinoma, a control group, and post-treatment patients. Their neural network achieves an overall accuracy of 82.09% for identifying nasopharyngeal carcinoma in control and post-treatment patients (Shu et al., 2021). Liu et al. review publications of Raman-based endoscopy for diagnostics of gastric cancer and list 10 papers that reported successful in-vivo tests of their endoscopic systems already in the first half of the 2010s. In these papers, the researchers used partial least-squares DA, LDA, and ANN models to classify Raman spectra.

The research group around Zherebtsov et al. introduced a spectroscopic approach that, in combination with AI classification, allows for real-time *in-vivo* cancer diagnostics (Zherebtsov et al., 2020). It is a fibre probe that implements two optical non-invasive modalities, the tissue endogenous fluorescence (TEF) and laser Doppler flowmetry (LDF), and is developed to allow the surgeons to fast and precisely determine the boundaries of a tumour directly during the surgeries. The authors tested their fibre probe under clinical conditions directly in patients with bile duct cancer and showed that it is extremely promising to target all four requirements of modern tumour diagnostics and treatment. Once the design of the probe is finalised and perfected, it will allow to significantly speed up the diagnostic process as the information about the tissue type and quality should be processed in real-time (I); contribute to patients safety as no tissue needs to be extracted for analysis (II); increase the precision of the diagnostic (III) and support the doctors' decision making (IV) via automated tissue type evaluation with AI approaches (Zherebtsov et al, 2020). Let us now take a closer look at the details of this approach.

Why did the authors decide to use such modalities as TEF and LDF for cancer diagnostics? TEF studies endogenous fluorophores, i.e. various substances in tissue cells that can fluoresce, and includes the tissue excitation with light at a certain wavelength and the subsequent collection of the fluorescence spectra (Chance & Williams, 1955; Chance, 1965; Lakowicz, 2006). Accordingly, TEF can be used for direct and fast tissue-state monitoring to facilitate minimally invasive surgery (Nazeer et al., 2018). Specifically in tumour diagnostics, TEF measurements are one of the most valuable methods to distinguish between healthy and malignant tissue (Conklin et al., 2009, Liu et al., 2011; Palmer et al., 2015; Awasthi et al., 2016; Wang et al., 2017; Kandurova et al., 2019). In particular, TEF emission of coenzymes such as reduced nicotinamide adenine dinucleotide (NADH) and oxidised flavin adenine dinucleotide (FAD) has been extensively investigated as these molecules play an important role in cells' metabolism and their TEF spectra can be used to track metabolic changes in cells and tissue (Koenig & Schneckenburger, 1994). NADH as an electron donor and FAD as an electron acceptor is involved in the production of adenosine triphosphate (ATP), a universal source of energy for many other biochemical reactions. The ratio between NADH and FAD in cells depends on different metabolic conditions, including pathological ones, and can be used to track oncogenesis (Croce & Bottiroli, 2014; Andersson-Engels et al., 2020; Druzhkova et al., 2016). In particular, the content of NADH and FAD is altered in the tumour as compared to the surrounding healthy tissue (Ju et al. 2020). Accordingly, there must be a difference between the TEF spectra obtained from cancerous and healthy tissue. Also, it is common knowledge that tumour cells grow and divide at a higher rate than healthy cells, which leads to changes in bioenergetic processes in cell mitochondria (Gulledge & Dewhirst, 1996). As a result, TEF spectra also encode information about the physiological or altered morpho-functional properties of the malignant cells (Ramanujam, 2000). To obtain TEF spectra, the authors include two excitation wavelengths in their fibre probe, in the near UV (365nm) and blue range (450nm). They collect the spectra pointwise from cancerous surrounding healthy material. The collection and subsequent labelling of the spectra are performed by the surgeon before they remove the tumour.

Now, let us consider why the authors chose to deploy laser Doppler flowmetry as a modality additional to the collection of TEF spectral from the tumorous. In tumour diagnostics, intravenous injection of special fluorescence dyes is often used to increase the fluorescence of the malignant cells (Sun et al., 2016; Zhao et al., 2018). However, as we discussed previously, the injection of chemical agents can compromise the patient's safety. The necessity of the agent injection can be circumvented by adding further optical modalities to analyse a wider range of interrelated parameters of tissue metabolism, blood circulation, and histological architecture (Volynskaya et al., 2008; Le et al., 2017). To do so, the authors deploy LDF. It allows for studying blood supply disorder in the tumour as compared to the blood supply in the surrounding healthy area (cf. Hoff et al., 2009). Blood microcirculation is an important parameter to sustain the vitality of the tissue as it facilitates the cells' metabolic activity as well as the level of tissue oxygenation and nutrition (Guven, et al., 2020). Ischemia, i.e. the lack of blood perfusion, though, affects tissue viability and leads to tissue degeneration or death. LDF is a highly sensitive technique to observe the rapid changes in capillary circulation (Almond & Wheatley, 1992). It evaluates the tissue blood perfusion and, thus, provides valuable information on the level and dynamics of the oscillatory changes in the blood flow in microcirculatory vessels. LDF is done via the illumination of the tissue with near-IR laser light and the subsequent registration of the Doppler shift that arises due to the light reflection at moving blood cells (Schabauer & Rooke, 1994; Rajan et al., 2009). That way, LDF provides continuous and real-time observation of the microvascular blood flow parameters including rhythmic oscillations that are caused by internal regulatory mechanisms in the vascular bed and externally through the work of the heart and lungs (Kvernmo, 1999; Kvandal, 2006). The authors use a laser at 1064 nm to collect LDF spectra.

The design of the fibre probe presented by the authors in Zherebtsov et al., 2020 allows for the parallel collection of the TEF and LDF spectra. The combination of TEF and LDF modalities does not only lead to insightful results on tissue vitality and tumour morphology but has a practical reason. Namely, blood is a well-known absorber of optical radiation in the visible range, especially at green light wavelengths (Bosschaart et al., 2014). The same range is of interest for the researchers when measuring TEF (Jacques, 2013; Dremin, 2016). Too much blood in the tissue distorts its spectral composition and introduces errors that might lead to the loss of valuable diagnostic information. The authors use LDF to compensate degradation of the TEF signal due to optical absorption (Zherebtsov et al., 2018). Also, LDF measurements allow the authors to track and adjust the pressure of the probe on the tissue, a variable that can dramatically influence the quality of the collected optical spectra (Obeid et al., 1990).

Although the authors report a quite sophisticated probe design that allows for a parallel collection of the TEF and LDF spectra for a subsequent ML-based evaluation, they also discuss several problems that can degrade the quality of the collected data. This first problem is the position of the probe. The authors report that it was not easy to properly determine the position of the optical probe in the bile duct, although X-ray control was applied. Being two-dimensional, X-ray control is not perfect in the depiction of the mutual position of internal structures, surgical instruments, and the optical probe. Nevertheless, this method is well established and, in some sense, optimal for the visualisation of large structures in routine surgical practice. Still, the capabilities of the method are already insufficient for more precise determination of the optical probe position in the duct, as well as of the presence of excess blood or bile. The volume of tissue covered by light from the probe is quite small (about 1 mm$^3$). Therefore, the slightest shift and the smallest amount of the aforementioned liquids distorts TEF spectra and LDF.

Another challenge constitutes data labelling. The authors discuss that a spectrum coming from a healthy tissue could have been falsely labelled as cancerous and vice versa. This is because it was tricky for the surgeon to match the position of the probe to the tissue type in some cases as the cancer cells grow and distribute in the bile duct wall randomly and not always visibly. Taking into account these limitations, the authors assume a high amount of initially mislabelled spectra which can degrade the performance of the ML algorithms.

Having several ML approaches (KNN, decision trees, SVM, random forest, and AdaBoost) to classify the TEF spectra into cancerous and healthy ones, the authors report that KNN and AdaBoost performed the best in this task. Thus, KNN showed following values: accuracy = 60%, sensitivity = 41%, and specificity = 73% for the 365 nm channel and accuracy = 62%, sensitivity = 46%, and specificity = 78% for the 450 nm channel. As for AdaBoost, they achieved: accuracy = 65%, sensitivity = 26%, and specificity = 87% for the 365 nm channel and accuracy = 64%, sensitivity = 37%, and specificity = 86% for the 450 nm channel. The evaluation of the LDF spectra showed a significant difference in the amplitudes of the blood flow oscillations due to breathing and the heartbeat when these oscillations pass healthy or cancerous tissue. Thus, the breath and heartbeat oscillations have higher amplitude when they are transmitted through healthy tissue as compared with the transmission through the tumour tissue, a result that authors use as a further tissue classification modality. Using logistic regression, they achieved: accuracy = 70%, sensitivity = 52%, specificity = 87% in the differentiation of the tissue type based on LDF. For both modalities, the ML metrics (accuracy, sensitivity, and specificity) show results that are far below 100% (which would denote a perfect classifier performance). Taking into account the problems of intra-surgery in-vivo data collection that the authors encountered, they consider these results highly promising.

Having proven the effectiveness of a combined modality of LDF and fluorescence spectroscopy as well as encountered and discussed the problems of in-vivo data collection, the authors significantly improved the technology and provided novel prototypes and even fully developed products. Thus, they report a fine needle optical probe for the excitation of NAD(P)H and FAD molecule at 375 nm to measure fluorescence lifetime and to collect diffuse reflectance spectra to predict hepatocellular carcinoma (HCC) in liver tissue. They first test this needle in a murine model and later in patients with assumed HCC. The measurements of diffuse reflectance spectra allow for tracking of tissue oxygen saturation and subsequent clear differentiation between the HCC tumour and the adjacent tissue. An LDA (linear discriminant analysis) model can automatically classify the liver tissue in mice with HCC, mice HCC, and human HCC when it is used to evaluate fluorescence lifetime and intensity parameters (Zherebtsov et al., 2022a, Zherebtsov et al., 2022b). Zharkikh et al. present a wearable device that allows for tracking and assessment of vascular function and oxidative metabolic status in pregnant women with type 1 diabetes mellitus. In this wireless device called "LAZMA-W2", the authors implement their previous achievements in LDF measurements and the collection of autofluorescence spectra of the skin. They successfully test it with seven pregnant diabetes-mellitus-affected women (Zharkikh, 2021).

The achievements presented in this section perfectly show that spectroscopic methods combined with the modality of laser Doppler flowmetry and ML approaches are already now sufficient for the implementation of low-invasive (in-vivo or wearable) fast methods that can significantly enhance the advances in tumour diagnostics. In the coming years, the designs will be finalised and ready to be used in clinical practice.

## 5. Conclusion

Cancer remains an urgent problem in modern society. Although the widely used diagnostic techniques such as computer tomography, magnetic resonance imaging, or biopsy analysis allow for accurate visualisation and determination of the size and type of a tumour. These techniques are either time-consuming and invasive (as the biopsy analysis) or relate to patients' safety concerns (as computer tomography or magnetic resonance imaging deploy chemical agents to enhance the imaging quality). Accordingly, scientists and engineers around the world are actively looking for novel approaches that might speed up the diagnostic process while minimising the risks to patients' safety, as well as help the doctors evaluate medical-record data for more precise diagnostics. These requirements let investigate and deploy artificial intelligence methods such as machine learning and neural networks for fast and automated patient data analysis and, thus, support the doctors in their decision-making. Novel spectroscopy-based techniques can make biopsy obsolete if combined with machine-learning and neural-network data analysis contributing to the development of fast and low- or non-invasive diagnostic methods. This review shows how AI methods have proven their effectiveness in the analysis of data generated by conventional diagnostic methods in oncology such as magnetic resonance imaging, the histopathological study of biopsy samples, and laboratory studies of biological liquids. Also, we reviewed highly promising spectroscopy-based methods for tissue imaging and analysis that might enhance the toolbox of effective and precise diagnostic methods. Spectroscopic methods allow for low-invasive, fast, label-free (i.e. without administration of chemical agents to the patient) tissue type imaging and analysis that can be done pre-, intra-operative, and post-operatively.

As an example of a spectroscopy-based approach, we saw a fibre optic probe that allows differentiating between cancerous and healthy tissue directly during surgeries. This probe excites NADH and FAD molecules in tissue cells by laser light in the near-UV and blue range to obtain tissue endogenous fluorescence spectra. An additional optical modality, namely a near-IR laser for laser Doppler flowmetry, allows for the compensation of the fluorescence signal degradation due to optical absorption by blood. AI-processed fluorescence and Doppler-flowmetry spectra show highly promising, although not yet perfect, results in the differentiation of tumours and healthy tissue in the hepatoduodenal area (Zherebtsov et al., 2020). Further design development might contribute to the enhancement of low-invasive, real-time tissue type assessment during surgeries.

When assessing the opportunities and success of artificial intelligence techniques for data processing in tumour diagnostics, it is important to stress its remarkable potential. Research conducted on the new methods of data acquisition and processing has highlighted several challenges that still need to be resolved. Thus, the limited amount of cancer-related data available for the development and training of machine learning and neural network models constitutes one of the biggest obstacles to the progress of AI-supported data evaluation. As for now, human estimation remains mostly more accurate in the determination of cancer, its stage and its type. However, the results presented in this review point to a steady improvement in the automatic processing of cancer-related data. Ultimately, it is to conclude that artificial intelligence technologies represent a truly powerful tool for cancer diagnostics, no matter if classical approaches (such as computer tomography or magnetic resonance imaging) are deployed or novel spectroscopy-based tools are used.

In addition to the approaches discussed in this review, it is worth mentioning that novel detectors including gas, solid-state, semiconductor, polymer-based, or plasmon-driven sensors are on the rise to collect data low- or even non-invasively for subsequent AI-driven evaluation to diagnose various types of cancer. As opposed to their classical quite bulky and expensive counterparts such as optical and mass spectrometers, CCD cameras and microscopes that require professional training to be used, novel sensors are portable and easier to deploy by semi-skilled people in the clinical or even private practice due to their miniaturised dimensions, more cost-effective production, operation costs, and simplified layout. Thus, various portable so-called electronic or e-nose implementations were proposed and successfully tested to diagnose lung cancer by the simultaneous non-invasive detection of organic compounds in human breath by arrays of different miniaturised sensors and a subsequent application of AI techniques to evaluate the data (Behera et al., 2019). In particular, plasmon-based sensors are promising versatile candidates for the analysis and data collection from different protein compounds, enzymes, antibodies, or nucleic acids (Pirzada & Altintas, 2020). To conclude, we see exciting developments in cancer diagnostics technologies that will soon contribute to the speed-up and precision increase of cancer diagnosis while minimising the risks for the patients.